\documentclass[11pt]{article}

\newcommand{\be}{\begin{equation}}
\newcommand{\ee}{\end{equation}}
\newcommand{\ba}{\begin{eqnarray}}
\newcommand{\ea}{\end{eqnarray}}
\newcommand{\rd}{\mathrm{d}}            
\newcommand{\re}{\mathrm{e}}            
\newcommand{\ri}{\mathrm{i}}
\newcommand{\rIm}{\mathrm{Im}}
\newcommand{\rRe}{\mathrm{Re}}

\newcommand{\te}[1]{\mbox{\boldmath $ #1 $}}
\newcommand{\tes}[1]{\mbox{\boldmath ${\it #1 }$}}
\newcommand{\bx}{\te{x}}
\newcommand{\bn}{\te{n}}

\newcommand{\bbe}{\te{e}}
\newcommand{\bu}{\te{u}}
\newcommand{\bv}{\te{v}}

\newcommand{\bF}{\te{F}}
\newcommand{\bU}{\te{U}}
\newcommand{\bE}{\te{E}}

\newcommand{\csch}{\,\mbox{csch}\,}

\newcommand\p{\ensuremath{\partial}}
\newcommand\sumn{\ensuremath{\sum_{n=0}^{{}\:\infty\:{}_\prime}}}
\newcommand\Pn{P_{n-\frac{1}{2}}}  
\newcommand\Qn{Q_{n-\frac{1}{2}}} 

\newcommand\etal{\mbox{\textit{et al.}}}

\newcommand\eg{\textit{e.g.}\ }
\newcommand\ie{\textit{i.e.}\ }

\usepackage{graphicx}

\begin{document}

\title{Surface tank-treading: propulsion of Purcell's toroidal swimmer}
\author {by Alex Leshansky$^1$ and Oded Kenneth$^2$
\\ \\ $^1$Department of Chemical Engineering, Technion, Haifa, 32000, Israel\\
$^2$Department of Physics, Technion, Haifa, 32000, Israel}
\maketitle

\begin{abstract}

In this work we address the ``smoking ring" propulsion technique, originally proposed by Purcell in \cite{purcell77}. We first consider self-locomotion of a doughnut-shaped swimmer powered by surface tank-treading. Different modes of surface motion are assumed and propulsion velocity and swimming efficiency are determined. The swimmer is propelled against the direction of its outer surface motion, the inner surface having very little affect. The simplest swimming mode corresponding to constant angular velocity, can achieve propulsion speeds of up to 66\% of the surface tank-treading velocity and swimming efficiency of up to 13\%. Higher efficiency is possible for more complicated modes powered by twirling of extensible surface.
A potential practical design of a swimmer motivated by Purcell's idea is proposed and demonstrated numerically. Lastly, the explicit solution is found for a two-dimensional swimmer composed of two counter-rotating disks, using complex variable techniques.

\end{abstract}

\section{Introduction}

Nano-technology is about the control of tiny objects. The benefits of tiny artificial swimmers for medicine, for example, could be enormous \cite{nanomed}. A tiny robot may swim through the arteries, digestive system, spinal canal, etc., transmit images and deliver microscopic payloads to parts of the body, or perform some kind of therapeutic action outside the reach of existing technologies.

Tiny swimmers, be they micro-organisms or microrobots live in a world dominated by friction \cite{purcell77}. Swimming at low Reynolds numbers is associated with energy dissipation and because of the absence of inertia, is often counter-intuitive. Micro-organisms had eons of evolution to optimize their locomotion and so are a natural source of inspiration and imitation in design of artificial microswimmers (e.g. \cite{actin}). Nevertheless, it is natural to ask whether the mechanisms used by micro-organisms are also the method of choice for artificial tiny devices. Microorganisms are different from artificial devices both in structure, time scales and functions, so a mechanism that is optimal for one may be inappropriate for the other. For example, useful robots would need to swim much faster than microorganisms. This implies that robots would be energy guzzlers compared with organisms which enjoy the additional luxury of swimming in a sea of nutrients. Autonomous artificial swimmers need also carry on ``battery" and/or ``engine", therefore, the issue of swimming  \emph{efficiency} becomes central. Swimming or \emph{hydrodynamic} efficiency rates different swimmers in terms of the power invested in swimming.

Swimming of micro-organisms has a rich history in applied mathematics \cite{childress} and several modes of natural locomotion, such as ``flexible oar" (\ie beating elastic tail) \cite{hancock53,WG98,WROG98}, the ``cork-screw" (i.e. rotating helical flagellum) \cite{hancock53,BW77,purcell97,berg,oren}, cilia \cite{cilia} or surface waves \cite{ESBM96,SS96} 
are quite well understood. Artificial biomimetic (e.g. nature-inspired) propellers, powered by either beating or rotating tails/filaments \cite{nature-blood,kosa,helical,robots,vomer} were recently fabricated and tested. As the above natural propulsion techniques are characterized by quite low hydrodynamic efficiency, it is both conceptually interesting, and technologically important to  understand what strategies may lead to effective propulsion in a setting dominated by viscous friction. Since tiny swimmers are a challenge to build while the theoretical framework is well understood it makes sense to put some effort into toying with mathematical models even if, at the moment, there may be no clear concept for their realization. Theoretical works on propulsion of artificial swimmers, that are not necessarily inspired by natural modes of locomotion, at low Reynolds numbers is a young enterprize and some mechanisms such as three-link Purcell's swimmer \cite{BKS03,TH07} three-linked-sphere swimmer \cite{najafi}, swimmer propelled by arbitrary non-retractable periodic shape strokes \cite{SW89,AGK04}, ``pushmepullyou" \cite{pmpy} and others were recently studied.
\begin{figure}[t]
\begin{center}
\includegraphics[width=3.2in]{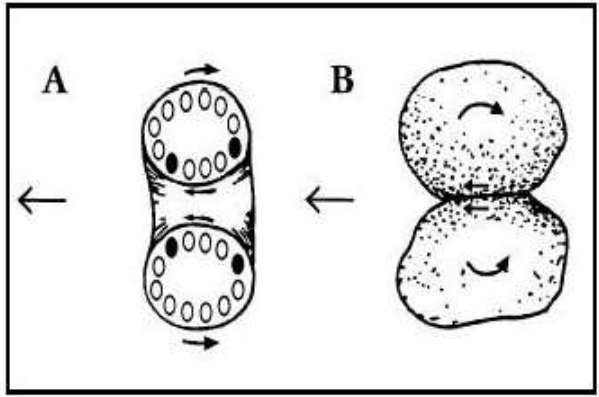}
\caption{Low-Reynolds-number-swimmers proposed by Purcell: (A) ``smoking ring" propulsion of a torus; (B) two cells stuck together and rolling on one another (redrawn from \cite{purcell77}) 
Arrows pointing to the left show the direction of propulsion, while smaller arrows at the swimmers' surface show the direction of the tank-treading motion.  \label{fig:Fig0}}
\end{center}
\end{figure}

A particularly promising class of strategies is where the motion is, in a sense, only apparent; where a shape moves with little or no motion of material particles. In our recent work \cite{LKGA07} we focused on locomotion powered by surface \emph{treadmilling}. In surface treadmilling the elongated microswimmer with length $\ell$ and thickness $d$ moves without a change of shape, by a tangential surface motion. Surface is generated at the front end and is consumed at the rear end. In contrast to actin and microtubules, the surface treadmilling does not rely on the exchange of material with the ambient fluid. (The swimmer needs, of course,  an inner mechanism to transfer material from its rear to its front which we did not account for in the model.) As the slenderness increases, the hydrodynamic disturbance created by the surface motion diminishes, i.e. the swimmer is propelled forward with the speed of boundary displacement backward, as the surface, except the near vicinity of the poles, remains stationary in the laboratory frame. The swimming movement is similar to the movement of tank treads, which propel the tank without the links dragging on the ground.
Therefore, the needle-shaped treadmill ($\varepsilon=d/\ell\ll 1$) is self-propelled throughout almost quiescent fluid yielding very low viscous dissipation and high swimming efficiency. One can not make treadmilling completely non-dissipative because there is always some remanent dissipation associated with the motion of the front and rear ends. It was demonstrated that the optimal ratio of power invested in dragging and swimming scales like $(\varepsilon \log{1/\varepsilon})^{-2}$ and can be made arbitrarily large.

In this works we consider a closely related mode of locomotion inspired by Purcell's idea \cite{purcell77} and powered by surface tank-treading: ``invaginating" torus (Fig.\ref{fig:Fig0}A) and two counter-rotating disks (similar to Fig.\ref{fig:Fig0}B). This technique, like the treadmilling, involves no change of shape, but, opposite to treadmilling, does preserve surface material. In this paper, we will address the hydrodynamic model of propulsion, swimming velocity, hydrodynamic efficiency, and other issues for three- and two-dimensional swimmer, respectively, We will also propose and numerically test a potential design of the microswimmer adopting this propulsion technique and composed of rotating microspheres.


 \section{Hydrodynamic model \label{sec:model}}

{\textit{General solution}--
In order to calculate the motion of the doughnut-shaped swimmer we introduce toroidal coordinates \cite{HB64} $(\xi,\eta)$ in the meridian plane  $(z,r)$ via the transformations
\be
z=c\frac{\sin{\eta}}{\cosh{\xi}-\cos{\eta}}\:,\qquad r=c\frac{\sinh{\xi}}{\cosh{\xi}-\cos{\eta}}\:, \label{eq:torcoor}
\ee
where $0\le\xi<\infty$, $0\le\eta<2\pi$ and $c>0$. It follows from (\ref{eq:torcoor}) that the curves $\xi=const$ form a family of nonintersecting coaxial circles with the centers in the plane $z=0$; the typical circle $\xi=\xi_0$ has its center at a distance $b=c\:\coth{\xi_0}$ from the origin and has a radius of $a=c\csch{\xi_0}$ (see Fig.\ref{fig:torus}). Upon rotation about $z$-axis these circles generate an eccentric family of toruses. The toroidal coordinates $(\xi,\eta,\varphi)$ form a right-handed system of orthogonal , curvilinear coordinates with metrical coefficients
$h_{\xi}=h_{\eta}=h=\frac{\cosh{\xi}-\cos{\eta}}{c},\, h_\varphi=\frac{\cosh{\xi}-\cos{\eta}}{c\,\sinh{\xi}}$.
\begin{figure}[tb]
\centering{\includegraphics[scale=0.7]{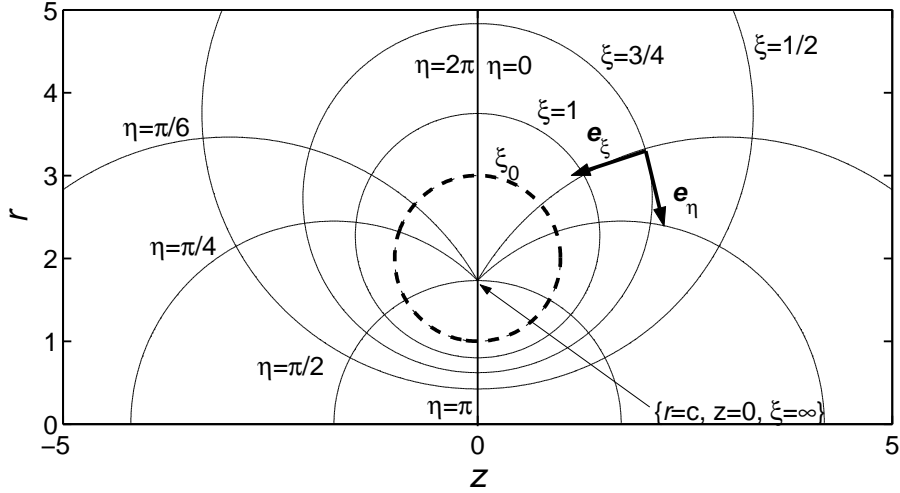}}
\caption{Toroidal coordinates in a meridian plane;  dashed line corresponds to a cross-section of a swimmer with $b/a=\cosh{\xi_0}=2$, while the torus  is obtained upon rotation around $z$-axis.   \label{fig:torus}}
\end{figure}

We assume that a steady axially symmetric creeping flow ($\mbox{Re}\ll 1$) has been established around the swimmer as a result of the tangential surface tank-treading (or ``smoking ring") motion with a uniform far-field velocity $U$ in the positive $z$-direction (\ie in the lab frame the swimmer is propelled with velocity $U$ in the negative $z$-direction). The flow is governed by the Stokes and continuity equations,
\be
\Delta \bv=\mu \: \mbox{grad}\:p\:, \qquad \mbox{div}\: \bv=0 \label{eq:stokes}\:,
\ee
respectively, with the boundary condition at the surface
\be
\bv=u(\eta)\:\bbe_\eta \qquad  \mbox{at} \qquad \xi=\xi_0\:. \label{eq:bc}
\ee

For the Stokes flow with axial symmetry the flow problem outlined in (\ref{eq:stokes}--\ref{eq:bc}) is reduced to determination of a stream function $\psi$ which satisfies in the region of the flow the equation
$L^2_{-1} \psi\equiv E^2(E^2\psi)=0$,
where
\[
L_{k}=\frac{\p^2}{\p z^2}+\frac{\p^2}{\p r^2}+\frac{k}{r} \frac{\p}{\p r}\,.
\]
The velocity components are readily obtained from $\psi$:
\be
v_\xi=-\frac{h}{r}\frac{\p \psi}{\p \eta}\:,\qquad v_\eta=\frac{h}{r}\frac{\p \psi}{\p \xi}
 \label{eq:vcomp}
\ee

Following \cite{PP60}, where the closely related problem of the uniform flow past an ``immobile" torus was considered, we seek a solution which is periodic in $\eta$ (of period $2\pi$) and satisfies the boundary conditions (\ref{eq:bc}) at $\xi=\xi_0$
It follows from (\ref{eq:vcomp}) and (\ref{eq:bc}) that
\be
\psi=\chi\:,\quad\frac{\p \psi}{\p \xi}=\frac{r}{h}\,u(\eta)\quad \mbox{at}\quad\xi=\xi_0\:.\label{eq:bc1}
\ee
and $\psi \rightarrow -\frac{1}{2}U r^2$ as $\rho \rightarrow \infty$, where $\rho=\sqrt{r^2+z^2}$ is the distance from the origin to the point in space. The constant $\chi$ should be found as a part of the solution. We follow \cite{PP60} and write
\be
\psi=-\frac{1}{2}U r^2+\psi'+\chi\,\psi_1\,\label{eq:psidecomp}
\ee
where $\psi'$ and $\psi_1$ give rise to zero velocity at infinity and satisfy (\ref{eq:bc1}) on $\xi=\xi_0$ via the conditions
\ba
\psi'=\frac{1}{2}\,U r^2\:, && \frac{\p\psi'}{\p \xi}=U r \frac{\p r}{\p \xi}+\frac{r}{h} u(\eta)\,,\label{eq:bcpsiprime} \\
\psi_1=1\:, && \frac{\p\psi_1}{\p \xi}=0\,, \label{eq:bcpsi1}
\ea
Note that for ``immobile" surface, \ie $u(\eta)=0$, the problem is identical to that considered in \cite{PP60} (\textit{glider} problem).
The solution for $\psi'$ can be constructed from $\psi^1$ and $\psi^3=r^{-2}\psi^{-1}$, where $\psi^k$ represent a generalized axially symmetric potential corresponding to the solution of $L_k\,\psi^k=0$ \cite{PP60,Weinstein53}, and yields
\be
\psi'=r^2 (s-t)^{1/2}\sumn\left[A_n\, s\, \Pn'(s)+B_n\, \Pn(s)\right]\cos{n\eta}=r^2 (s-t)^{1/2}\sumn W_n(s) \cos{n\eta}\,,\label{eq:psiprime}
\ee
where $s\equiv\cosh{\xi}$, $t\equiv\cos{\eta}$, $\sum\limits^{\;\;{\;}_\prime}$ indicates that the term for $n=0$ is to be multiplied by $1/2$ and $P_{n-1/2}'(s)={\rd\over\rd s} P_{n-1/2}(s)$, with $P_n$ being the Legendre function of the first kind \cite{AbramStegun}.

Substituting (\ref{eq:psiprime}) into the first equation in (\ref{eq:bcpsiprime}) and using the identity
\be
(s_0-t)^{-1/2}=\frac{2\sqrt{2}}{\pi}\sumn \Qn(s_0)\, \cos{n\eta}\:,\label{eq:identity1}
\ee
leads to the condition analogous to Eq. (3.15) in \cite{PP60}
\be
A_n\, s_0 \, \Pn'(s_0)+B_n\,\Pn(s_0)=U\frac{\sqrt{2}}{\pi}\, \Qn(s_0). \label{eq:condAB1}
\ee
Here $\mathrm{Q}_n$ stands for the Legendre function of the second kind \cite{AbramStegun} and $s_0\equiv\cosh\xi_0$.

Substitution of (\ref{eq:psiprime}) into the second equation in (\ref{eq:bcpsiprime}) and using (\ref{eq:condAB1}) yields after some algebra
\be
\sumn \frac{\rd W_n(s_0)}{\rd s_0}\,\cos{n\eta}=U\,\frac{\sqrt{2}}{\pi}\sumn \Qn'(s_0) \cos{n\eta}+ \csch^2\xi_0\,\frac{u(\eta)}{(s_0-t)^{1/2}}\,.\label{eq:bcpsiprime1}
\ee
In the particular case of constant tank-treading speed, \ie $u(\eta)=a\omega=v_s=\mbox{Const}$\footnote{the surface rotates clockwise (with respect to $\varphi$-axis) with the angular velocity $\te{\omega}=-\omega\:\te{e}_\varphi$}, the term $(s_0-t)^{-1/2}$ on the RHS of
(\ref{eq:bcpsiprime1}) can be expanded via (\ref{eq:identity1}) to yield
\be
A_n\, \frac{\rd }{\rd s_0} \left[s_0 \, \Pn'(s_0)\right]+B_n \,\Pn'(s_0)=U\,\frac{\sqrt{2}}{\pi}\,\Qn'(s_0)+ \frac{2\sqrt{2}}{\pi} v_s \csch^2\xi_0\,\Qn(s_0)\,.\label{eq:condAB2}
\ee

For $\psi_1$ we use the same representation as for $\psi'$ in (\ref{eq:psiprime}) with $A_n$ and $B_n$ replaced with $C_n/c^2$ and $D_n/c^2$, respectively,
\be
\psi_1=\frac{r^2}{c^2} (s-t)^{1/2} \sumn \left[ C_n\,s \, \Pn'(s)+D_n\, \Pn(s) \right]\cos{n\eta}\,. \label{eq:psi1}
\ee
The boundary conditions (\ref{eq:bcpsi1}) for $\psi_1$ are identical to those corresponding to the flow past a immobile torus, and,
therefore, the analysis is identical to that in \cite{PP60}, pp.85--86. Substituting (\ref{eq:psi1}) into the first equation in (\ref{eq:bcpsi1}) leads after some algebra to,
\be
C_n\, s_0 \, \Pn'(s_0)+D_n\,\Pn(s_0)=\frac{3}{\pi\sqrt{2}}\,\Qn^{-2}(s_0)\:,\label{eq:condCD1}
\ee
where $\Qn^{-2}(s_0)$ is the Legendre function of the 2nd kind and degree $-2$ \cite{AbramStegun}.
The second condition for $\psi_1$ in (\ref{eq:bcpsi1}) is treated in much the same as way as the analogous condition for $\psi'$ leading after some algebra to
\be
C_n\, \frac{\rd }{\rd s_0} \left[s_0 \, \Pn'(s_0)\right]+D_n \,\Pn'(s_0)=\frac{3}{\pi\sqrt{2}} \frac{\rd}{\rd s_0} \Qn^{-2}(s_0)\,. \label{eq:condCD2}
\ee
Therefore, (\ref{eq:condAB1}), (\ref{eq:condAB2}), (\ref{eq:condCD1}) and (\ref{eq:condCD2}) can be solved for the unknown coefficients $A_n\:,B_n\:,C_n$ and $D_n$ for all $n=0,1,2,\ldots$ in terms of the unknown propulsion velocity $U$ and the prescribed surface tank-treading speed $v_s$.

The value of the stream function at the boundary of the swimmer, $\chi$, is found from single-valuedeness of pressure $p$ upon integrating it over a closed contour (see \cite{PP60} for details)
\[
\oint_\mathcal{C} \frac{\partial p}{\partial \tau}\, \rd \tau= \oint_\mathcal{C} \frac{1}{r} \frac{\partial}{\partial n} (E^2 \psi)\: \rd \tau=0\:,
\]
where $\mathcal{C}$ is any closed contour in the meridian plane, which encloses the body profile, and $\bn$ and $\te{\tau}$ are the unit normal and tangent vector to the curve, respectively\footnote{The sense of $\te{\tau}$ is such that the system of $(\bn,\te{\tau},\te{e}_\varphi)$ is right-handed in this order}. Integration over the contour composed of the segment of the $z$-axis $-R \le z \le R$ and the semi-circle $\rho=R$ that joins its end points $(\pm R, 0)$ and letting $R \rightarrow \infty$ yields after some algebra
\be
\chi=-c^2 \left(\left.{\sumn\,A_n}\right/{\sumn\,C_n}\right)\:. \label{eq:chi}
\ee

The swimming speed $U$ should be determined from the requirement of force-free swimming. The net force exerted by the viscous liquid in Stokes flows (there is only one component $F_z$ due to axial symmetry) can be found from the asymptotics of the far-field as \cite{HB64}
\[
\frac{F_z}{8\pi\mu}=\lim_{\rho\rightarrow\infty}\frac{\rho(\psi-\psi_\infty)}{r^2}=
\lim_{\rho\rightarrow\infty}\frac{\rho(\psi'+\chi\psi_1)}{r^2}\:,
\]
where $\psi_\infty=-1/2\,r^2 U$ denotes the stream function corresponding to the uniform flow at infinity (in the frame fixed with the swimmer) and $\rho=\sqrt{r^2+z^2}=c\left({\frac{s+t}{s-t}}\right)^{1/2}$. Since $\rho=\infty$ correspond to the ``point" $\left\{\xi=0,\:\eta=0 \right\}$, the above expression for (dimensional) force can be evaluated as
\be
\frac{F_z}{8\pi\mu}=\frac{c\sqrt{2}}{2}\sumn \left[\left(A_n+\frac{\chi}{c^2} C_n\right)\left(n^2-\frac{1}{4}\right)+2\left(B_n+\frac{\chi}{c^2}D_n\right) \right]\:,\label{eq:force}
\ee
where we have made use of identities $\Pn'(1)=1/2(n^2-1/4)$, $\Pn(1)=1$.

Due to the linearity of (\ref{eq:condAB1},\ref{eq:condAB2}) the solution $\psi'$ can be decomposed as $\psi'=U\,\psi'_{(g)}+v_s\,\psi'_{(p)}$. The subscript 'g' stands for ``glider": a swimmer with an immobile (``frozen") surface ($v_s=0$) being dragged through the viscous liquid with the velocity $U=1$; and the subscript 'p' stands for ``pump": the surface of the torus undergoing continuous tank-treading motion with the prescribed tangential velocity $v_s=1$, while a finite force applied to the torus holds it in place, i.e. $U=0$. Therefore,
\be
\left(A_n, B_n\right)=U\,\left(A_n^{(g)}, B_n^{(g)}\right)+v_s\,\left(A_n^{(p)},B_n^{(p)}\right)
\:,\label{eq:ABdecomp}
\ee
and (\ref{eq:chi}) can be re-written as
\be
\chi=-c^2 \left(\left.{U\,\sumn\,A_n^{(g)}}+v_s\,\sumn\,A_n^{(p)}\right)\right/{\sumn\,C_n}\equiv -c^2\left(\alpha U+\beta v_s \right)
\:, \label{eq:chi1}
\ee
where $A_n^{(g)}$ and $A_n^{(p)}$ can be found by solving (\ref{eq:condAB1}) and (\ref{eq:condAB2}) for the ``glider" problem (with $U=1,\,v_s=0$) and the ``pump" problem ($U=0,\,v_s=1$), respectively, and $C_n,\,D_n$ from the complementary problem (\ref{eq:condCD1}--\ref{eq:condCD2}) of $\psi_1$. Further, substituting (\ref{eq:ABdecomp}) and (\ref{eq:chi1}) into (\ref{eq:force}) and equating the net force exerted on the freely suspended swimmer to zero yields the swimming speed $U$ as a linear function of the boundary velocity $v_s$:
\be
U=-\frac{\sum_n' \left[\left(A_n^{(p)}-\beta C_n\right)\left(n^2-\frac{1}{4}\right)+2\left(B_n^{(p)}-\beta D_n\right) \right]}{\sum_n'\left[\left(A_n^{(g)}-\alpha C_n\right) \left(n^2-\frac{1}{4}\right)+2\left(B_n^{(g)}-\alpha D_n\right)\right]}\times v_s\:.\label{eq:U}
\ee
With the swimming speed at hand, we can compute the value of the stream function at the swimmer surface, $\chi$, from (\ref{eq:chi1}) which completes the solution of the self-propulsion problem.

The expressions for the propulsion speed (\ref{eq:U}) and the viscous force (\ref{eq:force}) can be simplified by noting that 
\be
\left(n^2-\frac{1}{4}\right) C_n+2 D_n=\frac{3}{4}\,A^{(g)}_n\:, \label{eq:identity2}
\ee
that can be verified directly using the closed-form expressions for the coefficients\footnote{Note that $\pi$ is missing in  denominator of expressions for the coefficients $C_n$ and $D_n$ appearing in (3.32) and (3.33) of Pell and Payne's paper \cite{PP60}} corresponding to the glider problem derived in \cite{PP60}.

\section{Toroidal glider}

We numerically solve the infinite linear system of equations (truncated at some $n_{max}=L$) corresponding to both, the ``glider" and the ``pump" problems. Although explicit solution for the coefficient $A_n$--$D_n$ can be derived (as in \cite{PP60}), the resulting expressions are cumbersome, and involve integrals like $\int_1^{s_0} Q''_{n-1/2}(s)\,P_{n-1/2}(s) \rd s$ that require numerical integration, we have chosen therefore, to use here the numerical solution. The truncation level $L$ corresponds to the accuracy of $10^{-6}$ and varies between $L=45$ for $b/a=1.02$ and $L\le8$ for $b/a>2$.

To verify the numerical scheme we first addressed the ``glider" case, which is equivalent to the well-known problem of the axisymmetric flow past a rigid (i.e. ``immobile") torus that was studied rigorously \cite{PP60,DMOR76,MO77,Wakiya84} and approximately (in the spirit of slender-body theory) \cite{JW79,CH90}. The dimensionless drag force exerted on the torus (scaled with $6\pi \mu a U$) calculated as a function of aspect ratio $s_0=b/a$ is in excellent agreement with earlier results reported in \cite{MO77} for the range $1.3\le s_0 \le 4$ and with \cite{DMOR76} for the case of closed torus ($s_0=1$). For large values of $s_0$, the asymptotic form of the force can be obtained following \cite{MO77} by taking asymptotic forms for $P_{n-1/2}(s_0)$ and $Q_{n-1/2}(s_0)$, solving (\ref{eq:condAB1})--(\ref{eq:condAB2}) for $A^{(g)}_n$ and $B^{(g)}_n$ and substituting them into (\ref{eq:force}) together with the identity (\ref{eq:identity2}). Alternatively, using slender-body approximation \cite{JW79}, valid at $b/a\gg1$, the dimensionless force exerted on a thin torus reads
\be
\frac{F}{6\pi\mu a U} \sim \frac{4\pi s_0}{3(\log{8\,s_0+1/2})} \label{eq:force1}
\ee
It can be seen from Fig.\ref{fig:force} that the expression in (\ref{eq:force1}) is very accurate even at moderate elongation, and deviation from the exact result is less than 2\% already at $s_0=4$. As expected, in Stokes flows the drag force is controlled by the largest particle dimension and grows (sub-linearly) with $s_0$ as $F\sim s_0/\log{s_0}$ similarly to the needle-like slender body.
\begin{figure}[t]
\centering{\includegraphics{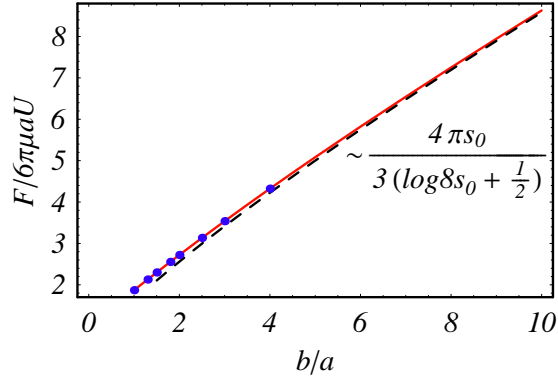}}
\caption{The dimensionless force exerted on the ``immobile" torus by the flow uniform at infinity vs. the aspect ratio $s_0=b/a$: current theory (solid line); slender-body theory \cite{JW79} (dashed line); numerical results from Table 1 in \cite{MO77} ($\bullet$). \label{fig:force}}
\end{figure}
The typical streamline pattern is depicted in Fig.\ref{fig:glider} for $s_0=2$.
\begin{figure}[t]
\centering{\includegraphics[height=8.5cm]{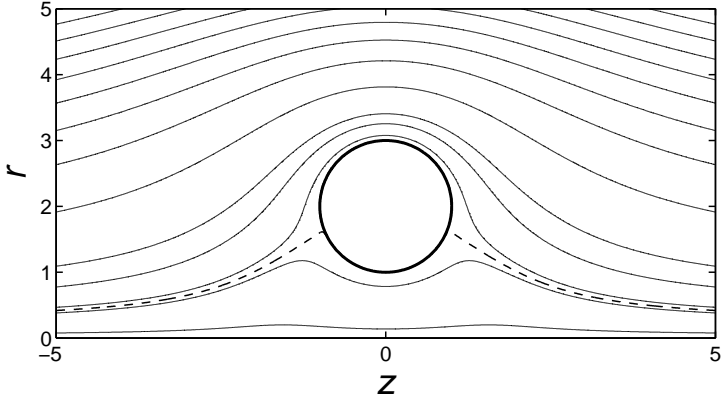}}
\caption{The streamline pattern corresponding to the flow around toroidal \emph{glider} in the meridian plane (the upper-half). The dashed line is the dividing streamline with $\psi=\chi$.  \label{fig:glider}}
\end{figure}

\section{Propulsion velocity of toroidal swimmer}

The dimensionless propulsion velocity $U/v_s$ of the toroidal \emph{swimmer} powered by the constant angular velocity of the surface is calculated via (\ref{eq:U}) as described in Sec.\ref{sec:model} and plotted vs. $b/a$ in Fig.\ref{fig:vel1} (solid line). The swimmer is propelled against the direction of its outer surface motion (\ie $U>0$) and the scaled propulsion velocity riches maximum, $U/v_s\simeq 0.665$ for a closed torus ($b/a\rightarrow 1$) and decays as slenderness ($b/a$) increases. The asymptotic form of the propulsion velocity at $b/a\gg 1$ can be derived by considering flow-field produced by line distribution of \emph{rotlets} $\bv_{rot}(\bx;s)=\Gamma \frac{\rd \te{c}(s)}{\rd s} \times (\bx-\te{c}(s))/(\bx-\te{c}(s))^3$ placed along the torus centerline $\te{c}(s)$ with arclength parameter $s$ and density $\Gamma=1/2 \omega a^2$ \cite{CH90}. Integrating  $\bv_{rot}$ over the length of the circle in the limit of large $s_0$ one obtains the net propulsion velocity \cite{KTS05}
\be
U \sim \frac{v_s}{2s_0} \left(\log{8 s_0}-\frac{1}{2}\right)\:.\label{eq:U1}
\ee
The agreement between the current theory and the asymptotic result (\ref{eq:U1}) (dashed line in Fig.\ref{fig:vel1}) is very good and it can readily be seen that the slender-body approximation provides an accurate estimate of the propulsion speed (within 1\% accuracy) starting from $s_0\simeq6$. The scaled propulsion speed decays slowly $\sim \log{s_0}/s_0$ for thin ring-like swimmers; for example, the swimmer with $b/a\sim 20$ (\eg twirling circular dsDNA ring \cite{KTS05}) would translate with the speed of $\sim$11.4\% of the tank-treading speed.
\begin{figure}[t]
\centering{\includegraphics{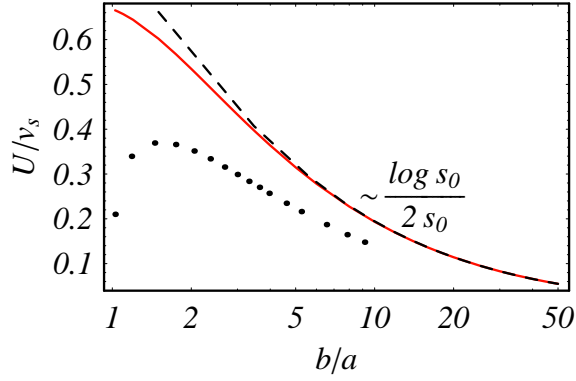}}
\caption{The scaled propulsion velocity of the toroidal \emph{swimmer}, $U/v_s$, powered by constant velocity tank-treading $u(\eta)=v_s=a\omega_\varphi=\mbox{Const}$ vs. the aspect ratio $s_0=b/a$: current theory (solid line); asymptotic result of \cite{KTS05}
(dashed line); swimmer made of closely packed (along the centerline) spheres rotating with constant angular velocity $\omega$ ($\bullet$).  \label{fig:vel1}}
\end{figure}
The typical streamline pattern corresponding to swimming powered by constant speed tank-treading at $b/a=2$ is depicted in Fig.\ref{fig:swimmer}.
\begin{figure}[t]
\centering{\includegraphics[height=8.5cm]{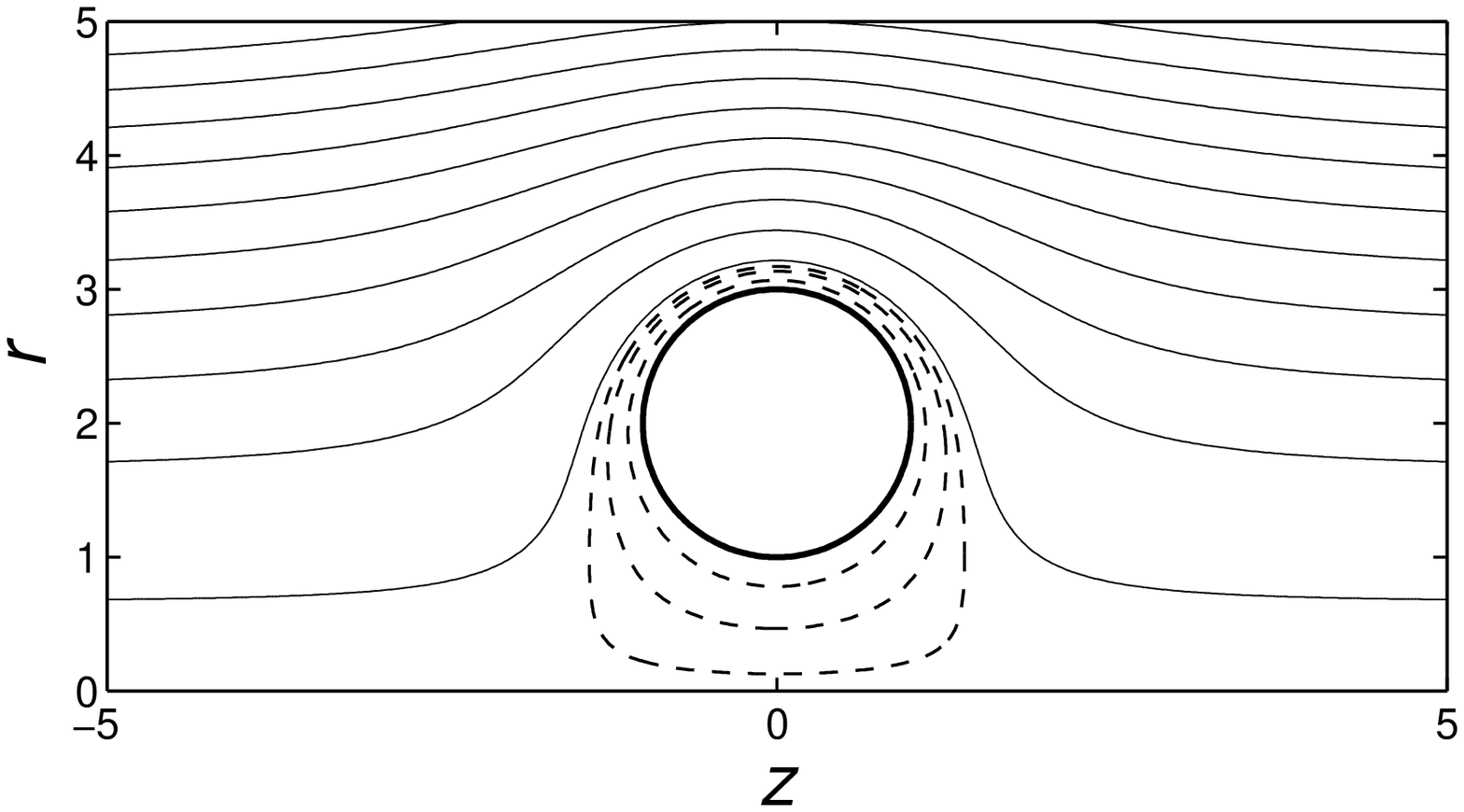}}
\caption{Flow around the toroidal \emph{swimmer} driven by constant-velocity-tank-treading motion of the surface. The dashed lines denote the closed streamlines in the near vicinity of the swimmer's surface. \label{fig:swimmer}}
\end{figure}

Stokes flows are rather counter-intuitive: the common intuition suggests that the swimmer should be propelled in the opposite direction as the hydrodynamic resistance to the flow inside the hole region is higher than outside. This expectation is invalid: the surface in the hole is undergoing tank-trading in the same direction with the same velocity everywhere dragging the fluid with it through the hole with relatively little resistance, while the outer portion of the surface is experiencing higher resistance as it is pushing against the stationary liquid at infinity

Two interesting questions are (i) whether the toroidal swimmer can perform any better in terms of the propulsion speed and (ii) what would be the propulsion speed for a biomimetic toroidal swimmer covered by highly deformable but almost inextensible membrane (biological membranes composed of lipid bilayer have a large modulus of dilation, i.e. they behave like two-dimensional nearly incompressible media). Since the swimmer moves in the direction of the inner surface motion, \ie it is propelled forward by the outer (``working") portion of the surface, one may expect that reduction/increase of the tank-treading velocity in the hole would not alter the locomotion speed-wise. In the case of a biomimetic surfac, the inextensibility  constrain imply that the smaller inner portion of the surface should move faster than the larger outer surface.
In this case, the surface velocity $\bu$ should satisfy $\nabla_s \cdot\bu=0$, where $\nabla_s=(\te{I}-\bn\bn)\cdot\nabla$ is the surface gradient operator. This condition in the particular case of an axisymmetric membrane becomes \cite{Pozrik03}
\be
\nabla_s\cdot\bu=\te{\tau}\cdot\frac{\partial\bu}{\partial s}+\frac{1}{r}\: \bu\cdot \te{e}_r=0\:,\label{eq:incompres}
\ee
where $\te{\tau}=\te{e}_\eta$ is a unit tangent vector to the surface in a meridian plane, and  $\te{e}_r$ is the unit vector in the direction of $r$. Obviously, the former case of constant tank-treading speed $\frac{\partial\bu}{\partial s}=0$ ($\bu=v_s \te{e}_\eta,\:v_s=\mbox{Const}$) is inapplicable for inextensible surfaces. Substituting $\bu=u(\eta)\te{e}_\eta$ and using the vector identity
$\te{e}_\eta=h(\frac{\partial r}{\partial \eta} \te{e}_r+\frac{\partial z}{\partial \eta} \te{e}_z)$ in (\ref{eq:incompres}) we arrive at
the 1st order differential equation for $u(\eta)$
\be
\nabla_s\cdot\bu=h\:\frac{\rd u}{\rd \eta}+\frac{1}{r} \, u \:(\te{e}_r\cdot\te{e}_\eta)=h\, \frac{\rd u}{\rd \eta}-\frac{\sin{\eta}}{c}\:u=0\:. \label{eq:incompres1}
\ee
The solution of this equation is straightforward and it defines $u(\eta)$ is a unique manner up to a multiplicative constant velocity $v_s^*$  that controls the magnitude of the tank-treading:
\be
u(\eta)=v_s^* (\cosh{\xi_0}-\cos{\eta})=v_s\:\frac{s_0-t}{s_0+1}\:.\label{eq:u-incompres}
\ee
We chose the constant $v_s^*$ equal to $v_s\:(\cosh{\xi_0}+1)^{-1}$ so the scaled tank-treading velocity $u/v_s$ riches a maximum value of 1 at $\eta=0$ for any $s_0$, \ie the surface velocity is scaled with the maximal boundary velocity $v_s$
which  for the incompressible membrane is attained at the inner surface of the torus.

The velocity distribution (\ref{eq:u-incompres}) can also be obtained by much simpler arguments. The conservation of the incompressible surface requires that the 2d flux $\oint_{\mathcal{C}} u_\eta\: r\rd\varphi=\mbox{Const}$, where the $\mathcal{C}$ is any closed circle at the torus surface at $\xi=\xi_0$ corresponding to some fixed value of $\eta$. Since the surface velocity $\bv=u(\eta)\:\bbe_\eta$ is constant along $\mathcal{C}$, the integration leads to $2\pi u(\eta) r=\mbox{Const}$, and thus $u(\eta)\sim 1/r$ and since at the swimmer surface $r=c\sinh{\xi_0}/(s_0-t)$ we readily arrive at (\ref{eq:u-incompres}).

The typical tank-treading velocity distribution (\ref{eq:u-incompres}) corresponding to incompressible surface is presented in Fig.\ref{fig:udist}\emph{a} for $s_0=1.1,\:2$ and $5$ vs. the polar angle $\theta$. $\theta$ is measured from $r$-axis about the center of the swimmer cross-section in the upper half-meridian plane, $(r=b,z=0)$ (see Fig.\ref{fig:torus}), so that $\cos{\theta}=(t s_0+1)/(s_0+t)$.
\begin{figure}[t]
\begin{tabular}{cc}
\includegraphics[height=4.5cm]{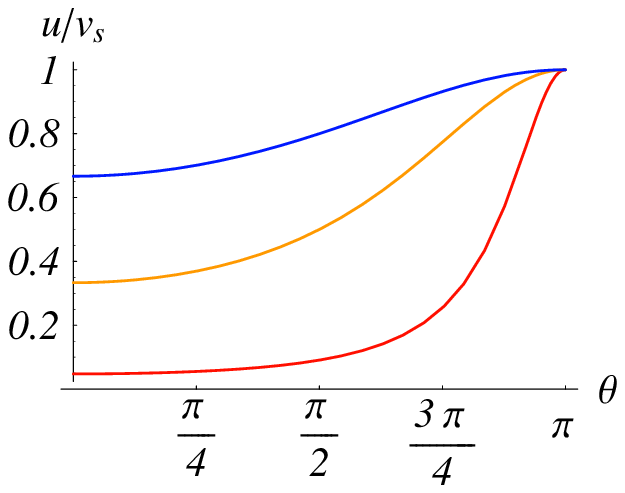} &
\includegraphics[height=4.5cm]{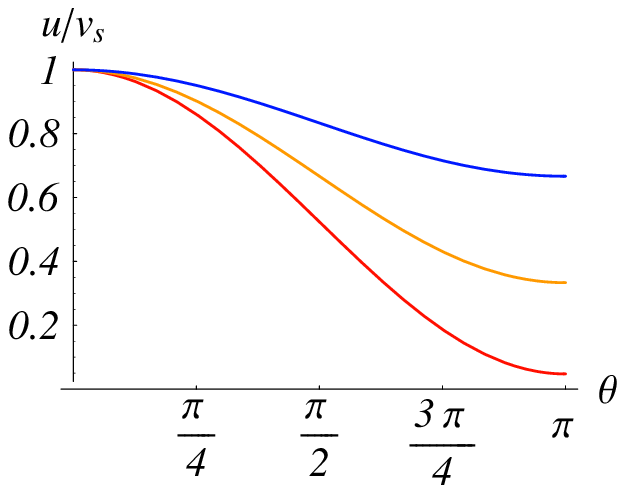} \\
(\emph{a}) & (\emph{b}) \\
\end{tabular}
\caption{Scaled tank-treading velocity $u/v_s$ vs. the the polar angle $\theta$ for  $s_0=1.1$ (red); $s_0=2$ (yellow); $s_0=5$ (blue). (\emph{a}) incompressible surface with velocity distribution (\ref{eq:u-incompres}); (\emph{b}) extensible elastic surface with velocity distribution
(\ref{eq:u-compres}. \label{fig:udist}}
\end{figure}
It is readily seen that for a nearly closed torus with $s_0=1.1$ the tank-treading velocity of the outer surface at $\eta\sim 0,\:2\pi$ almost drops to zero. Substitution of $u(\eta)$ from (\ref{eq:u-incompres}) into (\ref{eq:bcpsiprime1}) and some algebraic manipulations yield the  infinite set of equations analogous to (\ref{eq:condAB2}) for the coefficients $A_n$ and $B_n$
\ba
&& A_n\, \frac{\rd }{\rd s_0} \left[s_0 \, \Pn'(s_0)\right]+B_n \,\Pn'(s_0)=U\,\frac{\sqrt{2}}{\pi}\,\Qn'(s_0) \nonumber \\
&& \qquad + v_s \frac{2\sqrt{2}}{\pi} \frac{\csch^2\xi_0}{s_0+1}\:\left[-\alpha_n Q_{n-\frac{3}{2}}(s_0)+s_0 \Qn(s_0)-\beta_n Q_{n+\frac{1}{2}}(s_0)\right]\,,\label{eq:condAB21}
\ea
where $\alpha_n=0$ or $\frac{1}{2}$, and $\beta_n=1$ or $\frac{1}{2}$ according as $n=0$ or a positive integer, respectively. We solve the modified \emph{pump} problem (\ref{eq:condAB1}), (\ref{eq:condAB21}) in the same way as described in Sec.\ref{sec:model} (the \emph{glider} problem remains unaltered regardless of the particular tank-treading velocity) and plot the resulting scaled swimming speed $U/v_s$ vs. $s_0$ in Fig.\ref{fig:vel2} (yellow curve). The comparison with the propulsion speed of the analogous swimmer driven by constant-tank-treading motion (the red curve in Fig.\ref{fig:vel2}) suggests that the microrobot propelled by twirling inextensible surface is a rather lousy swimmer. Note that re-scaling the surface velocity in (\ref{eq:u-incompres}) with the speed of the outer surface, \ie writing $v_s^*=v_s\:(\cosh{\xi_0}-1)^{-1}$, yields propulsion speeds $U/v_s$ close to those corresponding to constant-twirling-velocity swimming (compare the  dashed yellow curve and the red curve in Fig.\ref{fig:vel2}). This supports the hypothesis that the motion of the outer surface is in control of propulsion, while tank-treading in the hole region affects only the swimming efficiency (see the next section).

%
\begin{figure}[t]
\centering{\includegraphics{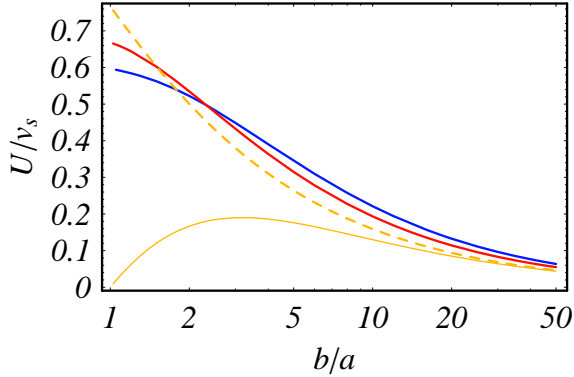}}
\caption{Dimensionless swimming speeds $U/v_s$ (scaled with the maximum tank-treading velocity of the membrane $v_s$) vs. the aspect ratio $s_0=b/a$:  constant tank-treading speed $u(\eta)=v_s$ (red); incompressible surface with velocity distribution (\ref{eq:u-incompres}) (yellow); extensible  surface with tank-treading velocity distribution (\ref{eq:u-compres}) (blue). The dashed yellow curve is corresponds to twirling of inextensible membrane re-scaled with the velocity of the outer boundary. \label{fig:vel2}}
\end{figure}

Now, one may also think of a swimmer propelled by tank-treading of highly extensible (metamorphic, \cite{nanomed}) elastic membrane, such that the outer surface moves faster than the inner surface. Let us assume a simple form of the velocity distribution satisfying this condition
\be
u(\eta)=\frac{v^*_s}{\cosh{\xi_0}-\cos{\eta}}=v_s\:\frac{s_0-1}{s_0-t}\:, \label{eq:u-compres}
\ee
where the non-dimensionalization is with respect to the maximum speed $v_s$ occuring in this case at $\eta=0$ on the outer surface. The typical velocity distribution (\ref{eq:u-compres}) is presented in Fig.\ref{fig:udist}\emph{b} for various elongations: $s_0=1.1,\:2$ and $5$.
Substituting $u(\eta)$ from (\ref{eq:u-compres}) into (\ref{eq:bcpsiprime1}) leads after some algebra to
\be
A_n\, \frac{\rd }{\rd s_0} \left[s_0 \, \Pn'(s_0)\right]+B_n \,\Pn'(s_0)=U\,\frac{\sqrt{2}}{\pi}\,\Qn'(s_0)-v_s \frac{4\sqrt{2}}{\pi}\csch^2\xi_0\,\Qn'(s_0)\:.\label{eq:condAB22}
\ee
The resulting propulsion speed of the torus powered by twirling of the extensible surface (\ref{eq:u-compres}) is provided in Fig.\ref{fig:vel2}
(blue curve). The scaled propulsion speed is similar to the other two results, represented by the red and the dashed yellow curves, respectively, in Fig.\ref{fig:vel2}) for nearly all values of $b/a$.

\section{Swimming efficiency}

The results in the previous section showed that the doughnut-shaped swimmer can propel itself with a speed of 66\% of the twirling velocity of its surface. Now let's determine how \emph{efficient} such swimmer is\footnote{By efficiency we mean only the \emph{hydrodynamic} efficiency; this does not include the work related to other forces, e.g. elastic forces within membrane}.

The power expended done by an arbitrary shaped organism and dissipated by viscosity in the viscous fluid is
\be
{\cal P}=-\int_S (\te{\sigma}\cdot\bn)\cdot \bu\: \rd S\:, \label{eq:dissp}
\ee
Eq.(\ref{eq:dissp}) can be written for the axisymmetric tank-treading of torus as ${\mathcal P}=\int_S \sigma_{\xi\eta}\,u_\eta\,\rd S$,
where $\sigma_{\xi\eta}$ in toroidal coordinates is given by
\[
\sigma_{\xi\eta}=h\left(\frac{\partial u_\eta}{\partial \xi}+ \frac{\partial u_\xi}{\partial \eta}\right)+\frac{1}{c}(u_\eta \sinh{\xi}+u_\xi \sin{\eta})\:.
\]
Using the fact that for swimmer propelled by purely tangential motion of its surface $u_\xi=\partial u_\xi/\partial\eta=0$, and applying (\ref{eq:vcomp}) to determine  $\partial u_\eta/\partial\xi$ from $\psi$ we integrate $\sigma_{\xi\eta} u_\eta$ in (\ref{eq:dissp}) numerically over $\eta$ from $0$ to $2\pi$ with $\rd S=2\pi c^2 \sinh{\xi_0}/(s_0-t)^2\: \rd \eta$.

The Lighthill's swimming efficiency $\delta$ is defined via
\be
\delta=\frac{\bF^{(g)}\te{\cdot}\:\bU}{\mathcal P}\:, \label{eq:delta}
\ee
where ${\cal P}$ is the rate of viscous dissipation in swimming with velocity $\bU$, and the expression in the denominator is the power expended by dragging the glider at velocity $\bU$ upon action of an external force $\bF^{(g)}$. For an axisymmetric swimmer $\bF^{(g)}\te{\cdot}\bU=\mathcal{R}_{FU}\,U^2$, where ${\cal R}_{FU}$ is the appropriate hydrodynamic resistance, this definition reduces to $\delta=\mathcal{R}_{FU}\,U^2/{\mathcal P}$. The efficiency $\delta$ is dimensionless and compares the particular swimming technique with dragging. The higher $\delta$ the more efficient the swimmer is. Using results from previous sections addressing the force exerted on the glider and the swimmer's propulsion velocity, we calculate $\delta$ for each one of the three modes: constant surface rotation velocity $u(\eta)=v_s=\omega a$, tank-treading of incompressible membrane (\ref{eq:u-incompres}) and twirling of extensible surface (\ref{eq:u-compres}). The resulting dependence $\delta(s_0)$ is provided in Fig.\ref{fig:delta}.
\begin{figure}[t]
\centering{\includegraphics{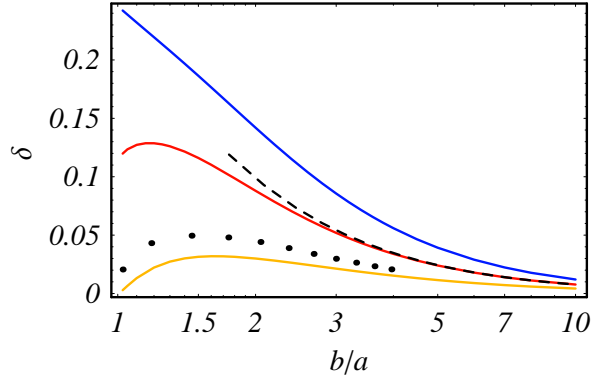}}
\caption{Swimming efficiency $\delta$ vs. the aspect ratio $s_0=b/a$:  constant angular velocity $u(\eta)=v_s=a\,\omega$ (red); incompressible surface with velocity distribution (\ref{eq:u-incompres}) (yellow); metamorphic surface with velocity distribution (\ref{eq:u-compres}) (blue); the dashed line correspond to the asymptotic result for a ``slender torus" (\ref{eq:delta1}); a swimmer made of closely-packed rotating balls ($\bullet$) as in Fig.\ref{fig:spheres}. \label{fig:delta}}
\end{figure}
As expected, the propulsion powered by the latter mode, \ie tank-treading by extensible surface with nonuniform velocity is superior over the other two cases. In the former case, $\delta \simeq 24$\% is achieved for a nearly closed torus $s_0 \rightarrow 1$. The constant twirling velocity swimming (the red curve in Fig.\ref{fig:delta}) yields maximal $\delta\simeq 13$\% for $s_0=1.15$, which is still better than other known propulsion techniques. The swimming efficiency of the torus powered by tank-treading of inextensible membrane (yellow curve in Fig.\ref{fig:delta}) is lower, as expected due to extra power invested in high-speed tank-treading of the inner surface in the hole, and has a peak $\delta \simeq 3.2$\% at $s_0=1.6$.

Swimming by extensible surface tank-treading is superior over the rotating helical flagellum \cite{purcell97}, beating flexible filament \cite{WG98}, the Percell's ``three-link swimmer" \cite{BKS03,TH07}, locomotion by virtue of shape strokes \cite{SW89,AGK04} and others. For comparison, swimming efficiency of undulating rod propagating plane waves of lateral displacement, has an optimum efficiency of $\delta=7.36\%$, while the most efficient swimmer made of two linked counter-rotating helixes of opposite handedness has a peak efficiency of $\delta=8.58\%$ \cite{hancock53}. Note that these values correspond to infinite \emph{unloaded} configurations. The swimmer propelled by undulating finite flexible filament attached to a passive cargo, as in experiments in \cite{nature-blood} is expected to do much worse than $\delta=7.36\%$; the experimentally determined hydrodynamic efficiency of bacterium \emph{e. coli} powered by rotating helical flagellum is $\approx 2$\% \cite{ecoli}. In our case, a useful cargo can be enclosed within the tank-treader without reducing the efficiency. The efficiency of spherical squirmers self-propelled by propagating surface waves along their surface (the mathematical model of cianobacteria \cite{ESBM96}) has the upper bound $\delta \le \frac{3}{4}$, while numerically calculated values of $\delta$ are usually less than 2\% \cite{SS96}; the maximum efficiency reached by the optimized three-link swimmer is only 1.3\% \cite{TH07}. Although tank-treading cannot, of course, beat the nearly frictionless treadmilling technique \cite{LKGA07}, where $\delta$ was shown to grow unbounded for a slender needle-like swimmer (as the backward motion of the inner surface and possible energy dissipation associated with it were not considered), the former mechanism is more attractive from practical point of view.

The asymptotic behavior of $\delta$ for slender torus, $s_0 \gg 1$, can be readily evaluated. To this end consider first the flow-field around a straight rod rotating about its axis, $\bv=\frac{a^2\omega}{r}\te{e}_{\varphi}$.
Substituting $\left.(\te{\sigma}\te{\cdot}\bn)\te{\cdot}\bv\right|_S=\left.(\frac{\partial v_\varphi}{\partial r}
-\frac{v_\varphi}{r})\,v_\varphi\right|_{r=a}=-2a\omega^2$ into (\ref{eq:dissp}) and multiplying by the surface area of the torus, equal to $4\pi^2ab$ we immediately arrive at the estimate $\mathcal{P}\sim 8 \pi^2\, \mu\, b\, v_s^2\:$.
This estimate is, actually, a lower bound for $\mathcal{P}$. Re-writing the dissipation integral as ${\cal P}=2\mu \int_V \bE\,\te{:}\,\bE \: \rd V$, where $\bE$ is the rate-of-strain tensor and expressing the product $\bE\te{:}\bE$ as $\sum \zeta_i \zeta_i+2 (\partial_i v_j) (\partial_j v_i)$, where $\te{\zeta}=\mbox{curl}\,\bv$ denotes vorticity, allows expressing ${\cal P}$ for microswimmers self-propelled by purely tangential motions\footnote{In the laboratory frame the velocity at the surface is a superposition of the translational velocity $\bU$ and purely tangential motions $\bu$} $\bu$ as \cite{SS96}
\be
{\cal P}=\mu\int_V\te{\zeta}^2 \rd V+2\mu \int_S\:\bu^2 \kappa_s\:\rd S\:. \label{eq:dissp2}
\ee
Here $V$ is the fluid volume surrounding the swimmer and $\kappa_s=-(\partial \te{\tau}/\partial s)\te{\cdot}\:\bn$ is the
curvature of the surface along the direction of the flow \cite{SS96} that is in case of a torus equal to $1/a$.
Since the creeping flow around infinite straight cylinder rotating around its axis is irrotational
(i.e. $\te{\zeta}=0$), the volume integral in (\ref{eq:dissp2}) vanishes as $b/a \rightarrow \infty$.
At finite $b/a$ we have, however, only an inequality
\be
\mathcal{P}\ge 2\mu \int_S\:\bu^2 \kappa_s\:\rd S=8 \pi^2\, \mu\, b\, v_s^2\:. \label{eq:Pasymp}
\ee
Comparing with the asymptotic form for the viscous force on a glider (\ref{eq:force1}) and with the propulsion speed
(\ref{eq:U1}), one can readily estimate $\delta$ from above as
\be
\delta \le \frac{\left(\log{8\,s_0}-\frac{1}{2}\right)^2}{4\,s_0^2 \left(\log{8\,s_0}+\frac{1}{2}\right)}
\sim\frac{\log{s_0}}{4 s_0^2}\:. \label{eq:delta1}
\ee
This asymptotic result is presented in Fig.\ref{fig:delta} as a dashed line and is in a very close agreement with the
exact result already for moderate $s_0$: the relative error between the asymptotic and the exact result is $\simeq 1\%$
for $s_0=5$.


\section{Microswimmer made of rotating spheres}
\begin{figure}[tb]
\centering{\includegraphics[scale=0.75]{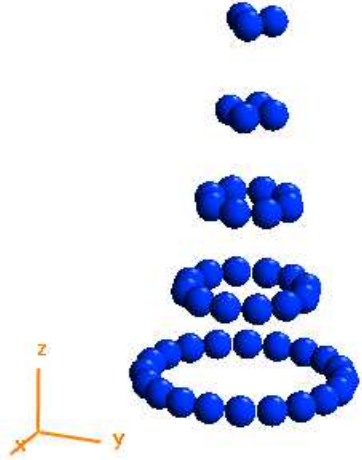}}
\caption{Illustration of necklace-like swimmers made of rotating spheres.  \label{fig:spheres}}
\end{figure}
In this section we propose a potential design for the swimmer propelled by the surface rotation. Imagine the necklace-like ring of radius $b$ in the $xy$ plane made of $N$ closely packed (separated by the distance $d=0.05 a$) spheres of radii $a$ (see Fig.\ref{fig:spheres}). In a cylindrical polar coordinate system $(z,r,\varphi)$, each sphere rotates at the constant velocity $\tes{\Omega}=-\Omega \te{e}_\varphi$ , which in the absence of external forces causes the necklace to swim, as described in the previous sections, along the normal to the plane of the ring in the negative $z$ direction. This concept may serve as a basis for potential experimental design of a micro(nano)robot based on Purcell's idea.

Calculations of the Stokes flow involving collection of spheres are rather standard today. We use the Multiple Expansion (ME) approach \cite{Filippov00} and construct the solution of the Stokes equations in terms of Lamb's spherical harmonics expansion. The no-slip conditions at the surface of all particles are enforced rigorously via the use of direct transformation between solid spherical harmonics centered at origins of different spheres. The ME method yields a system of $3\times N\times L\times (L+2)$ linear equations for the expansion coefficients and the accuracy of calculations is controlled by the number of spherical harmonics (i.e. truncation level), $L$,  retained in the series. In some situations, when the rigid particles in near contact approach each other, the hydrodynamic interaction involve strong lubrication forces ($\sim1/d$ singularity in  the viscous forces for ``squeezing" flow between two spheres) and the ME method requires a large number of spherical harmonics due to slow convergence. The ``near-field" (closed-form) lubrication ``patches" based on pair-wise hydrodynamic interaction \cite{KK91} can readily be introduced to greatly improve convergence. In our case, the hydrodynamic interaction between neighboring spheres is weaker as the distances between the spheres are kept fixed, and only involve shearing of the fluid in the thin gap produced by their rotation (the viscous force  $\sim\log{(1/d)}$ for ``sliding" motion between two spheres). $L \le 7$ was sufficient for all configuration with $N \le 28$ spheres to achieve an accuracy of less than 1\%.

The resulting propulsion speed scaled with $a\Omega$ is depicted in Fig.\ref{fig:vel1} ($\bullet$ symbols) for swimmers composed of from $N=2$ to $28$ rotating spheres. The maximum propulsion velocity of $U/a\Omega\simeq 0.37$ is obtained for ``quadrocycle" (\ie for $N=4$). Though the swimmer made of spheres moves slower than the rotating torus (see Fig.\ref{fig:vel1}), the propulsion speed is yet considerable.

For necklace-like swimmers made of $N$ identical microspheres as in Fig.\ref{fig:spheres} the rate-of-viscous-dissipation is calculated as ${\mathcal P}=\sum_j \mathcal{P}_j=N \mathcal{P}_{sph}$ where $\mathcal{P}_{sph}$ corresponds to dissipation rate due to a single sphere in the necklace. $j$th sphere in the necklace undergoes a rigid-body motion according to $\bv_j=\bU+\tes{\Omega}\times\te{r}_j$ with $\te{r}_j$ is the radius vector with origin at the particle center, $\bU$ is the swimming speed and $\tes{\Omega}$ is the angular velocity, and thus
\be
{\mathcal P}_{sph}=-\int_{S_j} (\te{\sigma}\cdot\bn)\cdot \bv\:\rd S=-\bU\te{\cdot}\bF_{sph}-\tes{\Omega}\te{\cdot}\te{L}_{sph}\:, \label{eq:dissp3}
\ee
where $\te{L}_{sph}=\int_{S_j} \te{r}_j\times (\te{\sigma\cdot}\bn)\:\rd S$ is a hydrodynamic torque exerted on each ($j$th) sphere of the swimmer. Since each sphere in the necklace is force-free, $\bF_{sph}=0$, and the swimmer is torque-free, $\sum_j (\te{L}_j+\te{R}_j \times \bF_j)=0$, ($\te{R}_j$ is the vector connecting the swimmer origin and the center of the $j$th sphere) the propulsion efficiency (\ref{eq:delta}) can be determined from
\be
\delta=\left|\frac{\bF^{(g)}_{sph}\te{\cdot}\:\bU}{\te{L}_{sph}\:\te{\cdot}\:\tes{\Omega}}\right|=\frac{{\mathcal R}_{FU}^{sph}\:U^2}{{\mathcal R}_{L\Omega}^{sph}\:\omega^2}=\frac{a^2\,{\cal R}_{FU}^{sph}}{{\cal R}^{sph}_{L\Omega}}\left(\frac{U}{v_s}\right)^2\:, \label{eq:delta2}
\ee
where $\mathcal{R}^{sph}_{FU}$ and $\mathcal{R}^{sph}_{L\Omega}$ are the appropriate hydrodynamic resistances functions (configuration dependent) for a single  sphere in the necklace. These functions can be readily calculated numerically via the ME method: $\mathcal{R}_{FU}^{sph} \sim \mu a$ is equal to the viscous force exerted on a single sphere in the \emph{glider} dragged through viscous liquid with velocity $U=1$, while $\mathcal{R}_{L\Omega}^{sph} \sim \mu a^3$ is equal to the torque $L$ exerted on a single sphere in the freely suspended \textit{swimmer} with all sphere rotating with the angular velocity $-\Omega\te{e}_\varphi$. The results are depicted in Fig.\ref{fig:delta} for $N=2$--$12$, while the maximum, $\delta \simeq 4.95\%$ is obtained, again, for ``quadrocycle".

\section{A two dimensional swimmer}

A natural two-dimensional analog of the toroidal swimmer consists of two discs counter-rotating at the same angular velocity $\omega$. As we show below this has a simple closed form solution. A realistic swimmer based on this description may be constructed by attaching two counter-rotating cylinders.

The two-dimensional Stokes equations are conveniently dealt with by employing the complex notations $v=v_x+\ri v_y$, $z=x+\ri y$ and $\partial={1\over2}(\partial_x-\ri\partial_y)$. In these notations the equations take the form $2\mu\,\bar{\partial}\partial v=\bar{\partial} p,\;\rRe(\partial v)=0$ and the general solution is given by $v=f+\bar{g}-z\bar{f'},\; p=-4\mu\, \rRe(f')$ with $f,g$ being arbitrary holomorphic functions (e.g. \cite{Langlois64}).

Finding $f,g$ explicitly is, in general, a difficult task. In order to solve the problem at hand of two counter-rotating discs of radius $a$ with their centers separated by a distance $2b$ we employed conformal mapping techniques. The details of these lengthy calculations are presented in the Appendix together with an explanation on how to extend it to a more general (nonuniform) boundary velocity on the discs. The final result, however, turns out to be fairly simple and is given by:
\ba
f(z)&=&{a^3v_s\over4b(b^2-a^2)}\left({z^2-b^2+a^2\over z^2+b^2-a^2}\right)\:, \label{fz}\\
g(z)&=&v_s\left\{-{a (2b^2-a^2)\over4b(b^2-a^2)}+{a(4b^2-a^2)\over2b (z^2+b^2-a^2)}-{a^3(b^2-a^2)\over b(z^2+b^2-a^2)^2}\right\}\:,\label{gz}
\ea
where $v_s=a\omega$ is the boundary velocity. Indeed, substituting $z=\pm \ri b+a\re^{\ri\phi}$ one readily finds $v=\pm \ri v_s \re^{\ri\phi}$. Since $f,g$ are regular everywhere outside the discs (including at infinity) one concludes that this is the correct flow solution.

At infinity we have $v(\infty)=-{ a\over 2b}v_s$ i.e. the fluid flows in the direction where the outer portion of the discs moves as in the 3D case (here the negative $x$-direction). The swimming speed, in the fluid rest frame, will then be
\be
\bU=+{ a\over 2b} v_s \te{e}_x\:. \label{eq:U2d}
\ee

The element of force acting on an infinitesimal segment may always be expressed as $dF=\ri p\, \rd z+(2\ri\mu\,\bar{\partial}v)\,\rd\bar{z}=2\ri\mu \,\rd (v-2f)$. Since $f$ in our solution is single-valued (no $\log$ term) one immediately concludes that the two discs exert no force on each other: $\oint \rd F=0$. This conclusion can also be derived from symmetry arguments by considering the effect of reversing the rotation direction of the discs.

It is clear that to maintain the rotation, the discs must exert on each other a non-zero torque. Straightforward calculation yields the torque on the disk
\[
L_{disk}=\rIm \oint\bar{z}\rd F=4\pi\mu a v_s\:.
\]
The dissipation occurring on each disc is given by
\[
\mathcal{P}_{disc}=-\rRe\oint\bar{v}\,\rd F=4\pi\mu v_s^2=L\, \omega\:.
\]
The total power on the two discs is $\mathcal{P}=2\mathcal{P}_{disc}$. We recall that in 2D the dragging problem admit no regular solution
within the Stokes approximation\footnote{This is known as the Stokes paradox and can be resolved by noting that far from the object the quadratic term $(v\cdot\nabla)v\simeq(U\cdot\nabla)v\sim \mu\Delta\bv$ cannot be neglected \cite{lamb}}. Thus defining the swimming efficiency as $\delta=(\bF^{(g)}\cdot\bU)/{\cal P}$ makes no sense in the 2D context in which $\bF^{(g)}$ is not defined. This may be considered as a mere issue of normalization. A natural measure for the efficiency of the 2D swimmer is given by the dimensionless ratio \cite{AGK04,LKGA07}
\[
\delta^\star={4\pi\mu U^2\over \mathcal{P}}={1\over8}\left({a\over b}\right)^2\:.
\]

\section{Discussion and conclusions}

In this work we considered a propulsion technique that involves no change in the swimmer's shape and is powered by surface ``tank-treading''. The underlying mechanism of propulsion is based on the difference in effective viscous friction to the rotation of ``inner" (\ie in the hole) and ``outer" torus surface. This idea for potential locomotion at low Reynolds number was originally proposed by Purcell \cite{purcell77} more than 30 years ago, but the adequate mathematical analysis (in terms of propulsion speed and efficiency) has not been provided until now. Exact solution to the Stokes problem for the doughnut-shaped swimmer was obtained here via expansion in toroidal (ring) harmonics; For the analogous 2D  swimmer the explicit closed form solution was found using complex variable techniques.

It is demonstrated that the doughnut-shaped swimmer outperforms its bio-inspired competitors -- swimmers powered by undulating or rotating flagella, both speed-wise and efficiency-wise. The swimmer moves in the direction of the ``inner" surface motion: the motion of the outer surface yields viscous thrust as it pushes against the motionless liquid far from the swimmer, -- the opposite from what common intuition would expect in inertia-dominated environment.(This intuition might be wrong even when inertial forces takes over: a vortex ring in the inviscid fluid is propelled in the  same direction as the toroidal swimmer in viscous liquid \cite{lamb}).
The swimming speed is dominated by the surface velocity on the outer side of the torus, the inner side having only small affect.
For nearly closed torus propelled by tank-treading of extensible (biomimetic) membrane it can get up to 66\% of the surface
velocity (for constant surface twirling velocity).
The hydrodynamic efficiency is more sensitive to the swimming mode. It can get up to $\sim$24\% for nonuniform surface twirling velocity and probably even higher for appropriate  modes.
As the hole radius, $b$, increases (for a fixed cross-section radius $a$), the propulsion velocity slowly diminishes as $\frac{U}{v_s}\sim \frac{\log{s_0}}{s_0}$, where $s_0=b/a$ and $v_s$ denotes the tank-treading velocity of the outer ``working" portion of surface. The swimming efficiency of the doughnut-shaped swimmer decays as $\delta\sim \frac{\log{s_0}}{s_0^2}$ for $s_0\gg1$. When we assume that the torus is powered by tank-treading of incompressible surface (bilipid membrane), the surface velocity of the inner (smaller) surface is significantly higher than that of the larger outer surface. Thus, the propulsion speed of such swimmer is considerably lower than that of the swimmers with extensible surface and it amounts to 19\% of the maximum tank-treading speed $v_s$ at $b/a=3.2$. The optimal hydrodynamic efficiency is $\delta\simeq3.19$\% at $b/a=1.6$, but it is yet comparable with efficiency of flagellated microorganisms \cite{ecoli}.

Since there is, probably, no simple route of building the doughnut-shaped tank-treader, we propose a practical design of the swimmer that mimic tank-treading propulsion mechanism. It is composed of $N$ rigid nearly touching spheres of radius $a$ assembled in a circle and rotating about the line of centers with common angular speed $\Omega$. The resulting hydrodynamic problem is resolved numerically using the Multiple Expansion Method. The ``quadrocycle" swimmer composed of four spheres is optimal in terms of both, speed, $U\simeq 0.37\,a\Omega$, and hydrodynamic efficiency, $\delta\simeq4.95$\%.

For the two-dimensional analogue of the invaginating torus, i.e. two close counter-rotating disks, we obtain an explicit solution for the hydrodynamic problem, the resulting swimming speed and hydrodynamic efficiency, being $U=\frac{v_s}{2\,s_0}$, and, $\delta=\frac{1}{8\,s_0^2}$. Therefore, as expected, a torus is a better  swimmer than its two-dimensional equivalent: the propulsion velocity for the 2D swimmer (\ref{eq:U2d}) and the asymptotic expression for slender torus at $s_0 \gg 1$ (\ref{eq:U1}) differ by a multiplicative factor of $\log{s_0}$.

\section*{Acknowledgments}
This work was supported by the Technion V.P.R. Fund. We thank J.~E. Avron and O. Raz for fruitful discussions.

\section*{Appendix}
\setcounter{equation}{0}
\renewcommand{\theequation}{A\arabic{equation}}

Here we consider the 2d flow around a pair of discs of radius $a$ centered at $(x,y)=(0,\pm b)$ and counter-rotating with angular velocity $\omega$. Our aim is to resolve the viscous (Stokes) flow around the disks. However most of the derivation can be extended to the case where an arbitrary velocity $w$ is prescribed as boundary condition on the two circles. We therefore present the derivation in a way that makes this extension obvious. We do, however, restrict ourselves to flow being symmetric with respect to the $x$-axis.

Using complex notations $z=x+\ri y,v=v_x+\ri v_y$ the general solution to the Stokes equations is expressible as $v=f+\bar{g}-z\bar{f'}\:,p=-4\mu\, \rRe(f')$ where $f,g$ are arbitrary analytic functions (not necessarily single-valued). Since we assume the flow is symmetric with respect to the $x$-axis we have $v(z^*)=v(z)^*$ which imply also $f(z^*)=f(z)^*,g(z^*)=g(z)^*$.

It is convenient to define new parameters $\alpha,\beta$ by $b=\beta{1+\alpha^2\over1-\alpha^2},a=\beta{2\alpha\over1-\alpha^2}$ (equivalently $\beta=\sqrt{b^2-a^2},\alpha={1\over a}(b-\sqrt{b^2-a^2})<1$). We also define a new complex coordinate $\zeta$ by $z=\ri \beta{1+\zeta\over1-\zeta}$.
The area outside the discs is mapped by this change of variables to the region $\alpha<|\zeta|<\alpha^{-1}$(in particular $z=\infty$ is mapped to $\zeta=1$). Standard bi-polar coordinates may be defined by $\xi+\ri\eta=\log\zeta$, however we will not make use of these. In terms of $\zeta$ the general solution takes the form $$v=f+\bar{g}+{1\over2}{1+\zeta\over1-\zeta}(1-\zeta^*)^2\bar{f}'$$

The functions  $f(\zeta),g(\zeta)$ may contain a $\log\zeta$ term in case they are multivalued. But apart from this, they are analytic functions in the annulus $\alpha<|\zeta|<\alpha^{-1}$ and may, therefore, be Laurent expanded as $f(\zeta)=\sum_{n=-\infty}^\infty a_n\zeta^n+\log(\ldots),\; g(\zeta)=\sum_{n=-\infty}^\infty b_n\zeta^n+\log(\ldots)$. The symmetry of the problem implies $a_{-n}=a_n^*,b_{-n}=b_n^*$. It is convenient to denote by $F,G$ the sums of positive powers appearing in the expansions of $f,g$. They are clearly analytic inside the disc $|\zeta|<\alpha^{-1}$. We also denote $\tilde{F}(x)=\sum_{n=1}^\infty a_n^*\,x^n=F(x^*)^*, \tilde{G}(x)=\sum_{n=1}^\infty b_n^*\, x^n=G(x^*)^*$. Thus, the functions $f,g$ can be expressed as:
\ba
f(\zeta)&=&F(\zeta)+\tilde{F}(1/\zeta)+\ri a_0\log\zeta   \\
g(\zeta)&=&G(\zeta)+\tilde{G}(1/\zeta)-\ri a_0\log\zeta+b_0\:. \label{fg}
\ea
The constant $b_0$ was arbitrarily appended to $g$, while the reflection symmetry implies it must be real. The constant $a_0$ must also be real in  order for the physical flow $v$ to be single valued. It may be remarked that the case under study (i.e. two rotating discs) posses also a symmetry  upon reflection about the $y$-axis which may be used to show that all the coefficients in the expansion of $f,g$ are real and in particular $a_0=0$.
However, for the sake of generality we shall not use this symmetry here.

Next the general solution $v(\zeta)$ should be matched with the prescribed boundary condition. Due to reflection symmetry it is enough to match  boundary condition on one circle. We, therefore, consider in the following only $|\zeta|=\alpha$ and use the relation
$\zeta^*=\alpha^2/\zeta$ to express $v$ as a function of $\zeta$ alone. We find
\ba\label{v=w}
v=F(\zeta)+\tilde{F}(1/\zeta)+\tilde{G}\left({\alpha^2\over\zeta}\right)+
G\left({\zeta\over\alpha^2}\right)+b_0+\ri a_0\:\log\alpha^2+   \\ \nonumber
+{1\over2}\,{1+\zeta\over1-\zeta}\left(1-{\alpha^2\over\zeta}\right)^2\left[
\tilde{F}'\left({\alpha^2\over\zeta}\right)-\left({\zeta\over\alpha^2}\right)^2 F'\left({\zeta\over\alpha^2}\right)
-\ri{a_0\over\alpha^2}\zeta\right]\:.
\ea
This should equal the given boundary condition $w(\zeta)$. For rotating discs we have $w(z)=\ri {v_s\over a}(z-\ri a)$ implying
$w(\zeta)=v_s\left({\alpha-\zeta/\alpha\over1-\zeta}\right)$, where $v_s=a\omega$.

We now operate on both sides of equation(\ref{v=w}) with: $\oint_{|\zeta|=\alpha}{d\zeta\over2\pi \ri}{1\over\zeta-\alpha\xi}$.
The analyticity of $F,G,\tilde{F},\tilde{G}$ inside the disc $|\zeta|<\alpha^{-1}$, allows using the residue theorem to compute the integrals  involving $F(\zeta),G({\zeta\over\alpha^2}),F'({\zeta\over\alpha^2})$. The integrals involving  $\tilde{F}({1\over\zeta}),\tilde{G}({\alpha^2\over\zeta}), \alpha\tilde{F}'({\alpha^2\over\zeta})$ may be calculated in a similar fashion by first changing the integration variable to $\zeta'=1/\zeta$.

There are two distinct cases to consider. In the case of $|\xi|<1$ one finds:
\ba\label{w1}
w_1(\xi)\equiv\oint_{|\zeta|=\alpha}{d\zeta\over2\pi \ri}{w(\zeta)\over\zeta-\alpha\xi}=
F(\alpha\xi)+G\left({\xi\over \alpha}\right)-{(\alpha-\xi)^2(1+\alpha\xi)\over2\alpha^2
(1-\alpha\xi)}\left(F'({\xi\over \alpha})+\ri{\alpha a_0\over \xi}\right) \\  \nonumber
+\ri{\alpha a_0\over2\xi}+b_0+\ri a_0\log \alpha^2-{1\over2}\tilde{F}'(0)
+{(1-\alpha^2)^2\over 1-\alpha\xi}\tilde{F}'(\alpha^2)\:.
\ea
For the rotating discs we have $w_1(\xi)={\alpha-\xi\over1-\alpha\xi}v_s$.

In the case of of $|\xi|>1$ the same integral becomes
\[
-\tilde{F}\left({1\over\alpha\xi}\right)-\tilde{G}\left({\alpha\over\xi}\right)-
{(\alpha-\xi)^2(1+\alpha\xi)\over2\xi^2(1-\alpha\xi)}\tilde{F}'\left({\alpha\over\xi}
\right)-{(1-\alpha^2)^2\over\alpha\xi-1}\tilde{F}'(\alpha^2)-{1\over2}\tilde{F}'(0)
+\ri{\alpha a_0\over2\xi}\:.
\]

We find it more convenient here to redefine $\xi$ as $\xi\rightarrow1/\xi^*$
so that the new $\xi$ again ranges inside the unit disc. It is also more convenient
to consider the complex conjugated expression. Then we can write
\ba\label{w2}
w_2(\xi)\equiv\left[\oint_{|\zeta|=\alpha}{d\zeta\over2\pi \ri}
{w(\zeta)\over\zeta-\alpha/\xi^*}\right]^* =
-F\left({\xi\over\alpha}\right)-G(\alpha\xi)+{1\over2}{\alpha+\xi\over \alpha-\xi}
(1-\alpha\xi)^2F'(\alpha\xi)           \\  \nonumber
-(1-\alpha^2)^2{\xi\over\alpha-\xi}F'(\alpha^2)-{1\over2}\ri\alpha a_0\xi-{1\over2}F'(0)
\ea
For the counter-rotating discs $w_2\equiv0$. More generally one may verify that if $w(\alpha \re^{\ri\theta})=\sum_n w_n e^{\ri n\theta}$ then
$w_1(\xi)=\sum_{n=0}^\infty w_n\xi^n,\;w_2(\xi)=-\sum_{n=1}^\infty w_{-n}^*\xi^n$.
Eliminating $G$ from eq(\ref{w1},\ref{w2}) we arrive at 
\ba\label{fff}
&{\cal W}(\xi)\equiv w_1(\xi\alpha)+w_2(\xi/\alpha)=F(\alpha^2\xi)-F\left({\xi\over\alpha^2}\right)
+{(1-\alpha^4)\xi(1-\xi)^2\over(\xi-\alpha^2)(\alpha^2\xi-1)}F'(\xi) &\\  \nonumber
&+(1-\alpha^2)^2\left({\xi\over\xi- \alpha^2}F'(\alpha^2)+{1\over1-\alpha^2\xi}\tilde{F}'(\alpha^2)\right)
+\ri a_0(1-\alpha^2){ 1-\xi\over1-\alpha^2\xi}-\rRe F'(0)+b_0+\ri a_0\log \alpha^2 &
\ea
For counter-rotating discs ${\cal W}(\xi)={\alpha(1-\xi)\over1-\alpha\xi}v_s$.

We obtained a non-local ODE for the function $F$. Note, however, that apart from the constant $F'(\alpha^2)=\left(\tilde{F}'(\alpha^2)\right)^*$
all occurrences of $F$ contain a multiple of $\xi$ as an argument. Thus, one may hope that expanding in powers of $\xi$ around the origin will give
useful relations that will allow
determining the coefficient of $F(x)=\sum a_nx^n$ in the Taylor expansion at least recursively. The first two relations obtained this way deserve to be presented explicitly:
$${\cal W}(0)=(1-\alpha^2)^2\tilde{F}'(\alpha^2)+b_0-Re F'(0)+\ri a_0\log\alpha^2-\ri a_0(1-\alpha^2)$$
$${\cal W}'(0)=(1-\alpha^2)^2\left[ \alpha^2\tilde{F}'(\alpha^2)-{1\over\alpha^2}F'(\alpha^2)\right]-\ri a_0\alpha(1-\alpha^2)^2$$
The first relation can be solved for $\tilde{F}'(\alpha^2)$ (thus also giving its complex conjugate $F'(\alpha^2)$). Substituting this into the second relation and separating the equation to its real and imaginary parts allows then to solve for $a_0$, $b_0$.
Using this to eliminate $F'(\alpha^2),\tilde{F}'(\alpha^2),a_0,b_0$ from Eq.(\ref{fff}) then leads to a considerably simpler equation. For the counter-rotating discs the equation takes the form
\be\label{eee}
(\alpha^2-\xi)(1-\alpha^2\xi)\left[ F(\xi\alpha^2)-F(\xi/\alpha^2)\right]+(1-\alpha^4)\xi(1-\xi)^2F'(\xi)=\alpha(1-\alpha^2)\,v_s\xi^2
\ee
In the more general case the non-homogeneous term on the r.h.s of (\ref{eee}) is replaced by a more complicated expression constructed  from the prescribed ${\cal W}$.

Substituting $F(x)=\sum_{n=1}^\infty a_nx^n$ in (\ref{eee}) one may recursively solve for all the coefficients $a_n$ in term of $a_1$. Indeed, this procedure appears to work also when the r.h.s is replaced by the expression corresponding to an arbitrary boundary condition $w$. For the special case of the rigid-body rotation with constant angular velocity the procedure leads to particularly simple solution for $a_n$'s that allows summation $F(x)=\sum a_n x^n$. One finds
$$
F(x)={\alpha^3U_0\over(1-\alpha^2)(1-\alpha^4)}x+C{x\over1-x}
$$
where $C$ is an arbitrary constant. Since $F(\zeta)$ is (by definition) analytic inside the disc $|\zeta|<\alpha^{-1}$ the constant must vanish, $C=0$. Using the other relations we have it is then a simple matter to find
$$
a_0=0,\;b_0={\alpha(1-\alpha^2+\alpha^4)\over(1-\alpha^2)(1-\alpha^4)}v_s
$$
$$
G(x)=-{\alpha x(2+2\alpha^4-2\alpha^2+\alpha^2x)\over2(1-\alpha^2)(1-\alpha^4)}v_s
$$
Using (\ref{fg}) and transforming back to $z$-coordinate we finally arrive at the Eqs.(\ref{fz})--(\ref{gz}).
\ba
f(z)&=&{a^3 v_s\over4b(b^2-a^2)}\left({z^2-b^2+a^2\over z^2+b^2-a^2}\right) \nonumber \\
g(z)&=&v_s \left\{-{a(2b^2-a^2)\over4b (b^2-a^2)}+{a(4b^2-a^2)\over2b(z^2+b^2-a^2)}-{a^3(b^2-a^2)\over b(z^2+b^2-a^2)^2}\right\} \nonumber
\ea

A comment on the case of arbitrary boundary condition $w$: the iterative solution of the generalized Eq.(\ref{eee}) depends on the chosen value of $a_1$. Since the correct value of $a_1$ is \emph{a priori} unknown the obtained $F(x)$ may be different from the correct solution up to a homogeneous solution $C{x\over1-x}$ with an arbitrary constant $C$. This issue may be resolved in two ways. The first (and most straightforward) is to notice that $C=\lim_{n\rightarrow\infty} a_n$ . One may use this to estimate $C$ and correct the solution accordingly. The other potential approach relies on the fact that although $C{x\over1-x}$ is a solution of the homogeneous equation (\ref{eee}) with $v_s=0$, it is not a solution of (\ref{fff}) with
${\cal W}=0$ (i.e in the passage from (\ref{fff}) to (\ref{eee}) some information was lost.)

\end{document}